\documentclass{article}

\usepackage[english]{babel}

\usepackage[letterpaper,top=2cm,bottom=2cm,left=3cm,right=3cm,marginparwidth=1.75cm]{geometry}

\usepackage[english]{babel}
\usepackage[T3,T1]{fontenc}
\usepackage{graphicx} 
\usepackage{cite}
\usepackage{soul}
\usepackage{amsmath}
\usepackage{amstext}
\usepackage{amsfonts}
\usepackage{amssymb}
\usepackage{amsthm}
\usepackage{color}
\usepackage{bm}
\usepackage{datetime}
\usepackage{xcolor}
\usepackage{multicol}
\usepackage{slashed}
\usepackage{extarrows}
\usepackage{vwcol}
\usepackage{stackengine}
\usepackage{tikz}
\usepackage{float}
\usepackage{upgreek}
\usepackage[colorlinks=true, allcolors=blue]{hyperref}
\usepackage{authblk}

\date{}
\title{A Sun-like star orbiting a boson star}
\author[1]{Alexandre M. Pombo}
\author[2]{Ippocratis D. Saltas}
\affil[1,2]{ \small CEICO, Institute of Physics of the Czech Academy of Sciences, Na Slovance 2, 182 21 Praha 8, Czechia}
\begin{document}
\maketitle

\begin{abstract}
The high-precision astrometric mission GAIA recently reported the remarkable discovery of a Sun-like star closely orbiting a dark object, with a semi-major axis and period of $1.4\, \rm{AU}$ and $187.8$ days respectively. While the plausible expectation for the central dark object is a black hole, the evolutionary mechanism leading to the formation of such a two-body system is highly challenging. Here, we challenge the scenario of a central black hole and show that the observed orbital dynamics can be explained under fairly general assumptions if the central dark object is a stable clump of bosonic particles of spin-0, or spin-1, known as a boson star. We further explain how future astrometric measurements of similar systems will provide an exciting opportunity to probe the fundamental nature of compact objects and test compact alternatives to black holes.
\end{abstract}
%
 
    \section{Introduction}\label{Intro}
%
The high-precision astrometric mission GAIA \cite{vallenari2022gaia,arenou2022gaia} recently released the observation of a Sun-like star in an orbit around a dark central object, reporting a measurement of the system's orbital parameters down to the $\sim 10^{-2}\, \%$ level. The total mass of the system is at $11.7\, M_{\odot}$, the distance between the two objects is $1.4 \, \rm{AU}$, while the orbital period is amongst the highest ever observed for such a system, at about $187.8$ days \cite{el2022sun}. This benchmark observation is only the beginning, as GAIA promises to observe a large number of similar binary systems in the future towards a deeper understanding of their evolutionary dynamics. 
\\

The luminous star orbiting the dark object is compatible with a typical G-dwarf star of a mass $0.93 \, M_{\odot}$ at solar metallicity. The nature of the central black object, however, appears rather challenging to explain\footnote{We note that there have not been any X-ray or radio-wave observations associated to the central object that could help to identify it with a black hole.}. If the central object were a system of non-luminous baryonic compact objects such as neutron stars, it would probably be unstable.
On the other hand, the scenario of a central black hole requires unreasonable amount of fine-tuning within the usual evolutionary channels. In particular, if the system is expected to have formed as a binary in isolation, a common envelope formation scenario is rather unlikely, given the system's arrangement. This requires an extreme, and possibly unphysical, tuning of the relevant parameters of the evolutionary channel under consideration. Moreover, formation within a globular cluster is also improbable given the geometrical characteristics of the observed orbit. Other evolutionary channels such as formation without a common envelope or via a hierarchical triple also seem unlikely for similar reasons. For a detailed discussion of these issues we refer to \cite{el2022sun}. 
\\

In this work we will attempt to explain GAIAs's observation assuming the central object to be a compact, self-gravitating configuration of bosonic particles, known as a boson star. {\it Our motivation is twofold:}
{\bf i)} Precision astrometric observations of binary systems allow dynamical tests of gravity at astrophysical scales, and in particular, tests of black holes, as well as of horizon-less alternatives such as boson stars. Boson stars arise in theories where the role of dark matter is played by some massive field of bosonic nature. 
{\bf ii)} In addition, future high-precision observations can yield data that can help to distinguish between black holes from exotic alternatives in binary systems. It is, therefore, tantamount to test the geometrical differences of orbiting stars around possible dark objects so that one can gain insights not only on the evolutionary dynamics of binaries, but also on the properties of fundamental exotic particles which may constitute the elusive dark matter. 
\\

 
A boson star consists of a self-gravitating object made out of (exotic) boson particles in the form of a Bose-Einstein condensate. These may be described as everywhere regular lumps, \textit{i.e.} self-gravitating solitons, of some yet undetected bosonic field ~\cite{kaup1968klein,ruffini1969systems, brito2016proca, colpi1986boson,lynn1989q,schunck1998rotating,yoshida1997rotating,astefanesei2003boson,schunck2003general,liebling2017dynamical,grandclement2014models,herdeiro2017asymptotically,alcubierre2022extreme,herdeiro2019asymptotically,guerra2019axion,delgado2020rotating,minamitsuji2018vector,herdeiro2021multipolar,herdeiro2020asymptotically,herdeiro2023proca}. Boson stars are expected to form at astrophysical scales via collapse under their own gravity while dispersing energy through gravitational cooling \cite{seidel1994formation,guzman2006gravitational}, leading to a non-luminous lump of bosonic matter.  A key difference with black holes is the lack of an event horizon. 
\\

While boson stars with a generic bosonic field seem to be possible, only two fundamental fields have been studied in the literature: A bosonic matter of spin-0 ($s = 0$, scalar particles), and a spin-1 matter ($s = 1$, vector particles). For both cases, we will assume that the bosonic matter is coupled minimally to the spacetime curvature, however, we will allow for self-interactions of the bosonic fields. The values of the interaction couplings , including the mass of the boson, determine the effective mass and radius of the boson star. As we will show, for a large part of the parameter space of the bosonic interactions, we can construct boson star configurations which explain the orbital characteristics as measured by GAIA without the need of a central black hole. At the same time, we discuss how the future measurement of orbital precession of similar binary systems would allow to distinguish the nature of the central dark object. 
\\

{\it The paper is structured as follows:} In Section \ref{sec:theory}, we explain the main equations and numerical solutions for the construction of the boson stars which will play the role of the central dark object. In Section \ref{sec:orbits} we compute and discuss the orbital trajectories of the star around the boson star under different assumptions. Our results for the orbital dynamics of the two-body system are presented in Sections \ref{sec:CaseI} and \ref{sec:CaseII}. We summarise in Section \ref{sec:summary}. An overview of our results can be found in Figures \ref{F1}, \ref{fig:Orbits-scalar} and \ref{fig:Orbits-vector}.
\\

Unless otherwise stated, throughout the text we define the physical units as:  $c = 63241$ $\rm{AU} \cdot \rm{yr}^{-1}$, $G \simeq 39.748$ $\rm{AU}^3 \cdot M_{\odot}^{-1} \cdot \rm{yr}^{-2}$, $\rm{M_{P}} \simeq 1.094 \cdot 10^{-38} \ M_\odot$. For the numerical construction of the boson stars, we use geometrised units where $G=1=c=\hbar$. We are solely interested in spherical symmetry and the metric matter functions are only radially dependent. For notation simplicity, after being first introduced, the functions' radial dependence is omitted, \textit{e.g.} $X(r)\equiv X$, and  $X' \equiv dX/dr$.

%
    \section{Modelling the dark object}\label{sec:theory}
\subsection{Field equations for boson stars}

    The action that describes a self-gravitating bosonic field $\Psi$, minimally coupled to gravity reads
        \begin{equation}
	   \mathcal{S}=\int d^4 x \sqrt{-g} \left[ \frac{R}{16 \pi G}+\mathcal{L}_s(\Psi) \right],
	\end{equation}
    where $R \equiv R\, (g_{\mu \nu})$ is the Ricci scalar of the spacetime described by the metric $g_{\mu \nu}$,  $g$ is the  metric's determinant, and $G$ stands for the bare Newton's constant. Let us consider a complex spin-$0$ (represented by a scalar, $\Phi (r)$), and a complex spin-$1$ (represented by a vector, $A_\alpha (r)$). The respective lagrangian  $\mathcal{L}_s$ reads as
	\begin{equation}
	   \mathcal{L}_{0} = -\frac{1}{2}g^{\alpha \beta} \big(\partial_{\alpha}\bar{\Phi}  \partial_{\beta}\Phi + \partial_{\beta}\bar{\Phi} \partial_{\alpha} \Phi \big) + U_i(|\Phi |^2)\ , \qquad \qquad \mathcal{L}_{ 1} = -\frac{1}{4} F_{\alpha \beta}\bar{F}^{\alpha \beta}-V(\textbf{A}^2)\ ,
	\end{equation}
    with $\textbf{A}^2\equiv A_\alpha \bar{A}^\alpha$ and an overbar denoting complex conjugation. For both types of fields we consider a kinetic term and an interaction potential. The fields enjoy a $\textbf{U}(1)$ symmetry associated to a conserved Noether charge, which energetically stabilizes the solution. Variation of the action with respect to the metric and matter fields leads to the following sets of field equations\footnote{We remind that in this section we set $G=1$.}
        \begin{align}
	   & G_{\alpha \beta} = { 4\pi} \Big[\partial_{\alpha} \bar{\Phi} \partial_{\beta} \Phi +  \partial_{\beta}\bar{\Phi}  \partial_{\alpha} \Phi  -g_{\alpha \beta} \mathcal{L}_0 \Big]\ , \qquad \Box \Phi = \hat{U_i}\cdot \Phi \ ,\\
	   & G_{\alpha \beta} ={4\pi  }\left[\frac{1}{2} \big(F_{\alpha \delta} \bar{F}_{\beta \gamma} +\bar{F}_{\alpha \delta} F_{\beta \gamma} \big) g^{\delta \gamma}+\big( A_\alpha \bar{A}_\beta +\bar{A}_\alpha A_\beta - g_{\alpha \beta}\mathcal{L}_1 \big)\hat{V}\right] \ , \qquad\frac{1}{2} \nabla_\alpha F^{\alpha \beta} = \hat{V} \cdot A^\beta\ ,
	\end{align}
where $G_{\alpha \beta}$ is the Einstein's tensor, $\Box$  $(\nabla)$ the covariant d'Alembertian (derivative) operator, $\hat{U_i}\equiv d U_i/d |\Phi |^2$ and $\hat{V}\equiv dV /d \textbf{A}^2$. For the metric ansatz, let us consider a widely used choice compatible with spherically symmetric configurations 
        \begin{equation}
         k\, ds^2 = -\sigma (r) ^2 N(r)\, c^2 dt^2 + \frac{dr^2}{N(r)}+ r^2 \big(d\theta ^2 +\sin ^2\theta d\varphi ^2\big)\ ,\qquad N(r)=1-\frac{2\, m(r)\, G}{r c^2}\ ,
         \label{E5}
        \end{equation}
    with $m(r)$ the Misner-Sharp mass function~\cite{misner1964relativistic} and $\sigma (r)$ a free metric function. We use $k=1=c=G$ unless otherwise stated. We are interested in asymptotically flat, stationary and fundamental state solutions for the self-gravitating matter fields. For this purpose, we seek solutions under a harmonic ansatz as 
	\begin{equation}
	 \Phi (r,t) = \phi (r)\, e^{-i \omega t}\ ,\qquad \qquad A_\mu =\big[ f(r)\, dt + \textit{i} g(r)\, dr \big] e^{-i\omega t}\ ,
         \label{E6}
	\end{equation}
    where $\phi (r) $ is the scalar field's amplitude, and $f(r)$ and $g(r)$ are two real potentials that define the Proca (vector) field's ansatz. In both cases, $\omega$ is the field's frequency and is strictly real for stationary solutions. In the vector case, the field equation implies the Lorenz condition $\nabla _\alpha (\hat{V}A^\alpha) =0$, which is a dynamical condition, rather than a gauge choice.
    For the interaction potential, let us consider two well-motivated cases for the scalar field ($U_i$), and one for the vector field ($V$).
  
        {\bf Vector-field potential ($V$)}: For the vector model, we will focus on the simple case of a massive vector  \cite{brito2016proca,herdeiro2017asymptotically}
           \begin{align}
	      V = \frac{\mu _P ^2}{2} \textbf{A}^2\ ,
            \label{E7}
	    \end{align}	 
       with $\mu _P$ the vector's mass. Stars built out of vector bosons with a quartic self-interaction, \textit{i.e.} an interaction of the form $\sim \lambda \, \textbf{A}^4$, are also possible and were studied in \cite{minamitsuji2018vector,herdeiro2020asymptotically,herdeiro2021imitation}. However, the work of \cite{herdeiro2021imitation} showed that the allowed values of the vector's self-interaction have a negligible impact, while recent works have shown that self-interacting vector fields are prone to ghost instabilities \cite{clough2022ghost,clough2022problem}. For the potential \eqref{E7} the equations boil down to a set of four, first-order field equations, two for the metric and two for the matter field respectively,
\begin{align}
& m' = {4\pi}r^2 
	\left[ \frac{(f'-\omega g)^2}{2\sigma ^2} + \mu _P ^2 \Big(\frac{f^2}{2 N \sigma^2}+N\Big)\right] \ , \qquad \qquad	\sigma' = {8\pi} r \sigma \mu _P ^2 \left[ g^2+ \frac{f^2}{N^2 \sigma ^2}\right]\ ,\nonumber \\
& f' = \omega g-2\frac{g \sigma ^2 N}{\omega} \mu _P ^2\ , \qquad \qquad \left[\frac{r^2\big(\omega g - f'\big)}{\sigma}\right]'+\frac{2 r^2 f}{N \sigma}\mu_P ^2 =0\ . 	
          \label{E8}
\end{align}
     {\bf Scalar-field potential ($U_i$)}: Many boson star models based on scalar fields have been considered over the years. We refer to \textit{e.g.} \cite{schunck2003general} for a review. Here, we shall divide our analysis into two distinct cases. \textbf{i)} The first case is a potential up to a quartic self-interaction which corresponds to a renormalisable potential with even powers of the scalar field ~\cite{colpi1986boson}, and \textbf{ii)} an axion-like potential motivated by models of the QCD axion ~\cite{guerra2019axion,delgado2020rotating}. They read as 
        \begin{align}
          &U_{\rm{self}}= \frac{\mu _S ^2}{2}\, \phi ^2 +  \lambda\,\phi ^4\ , \qquad   \qquad U_{\rm{axion}} = \frac{2\,\mu _S ^2\, f_\alpha ^2}{\hbar B}\Bigg[1-\sqrt{1-4B   \sin ^2 \Big(\frac{\phi\, \sqrt{\hbar}}{2\, f_\alpha}\Big)}\, \Bigg]\ .
        \label{E9}
	\end{align}	 
     Where $\mu _S$ is the scalar field's mass, while $\lambda$ and $f_\alpha$ are the corresponding coupling strengths, and $B=\frac{z}{1+z^2}\approx 0.22$ with $z=\frac{m_u}{m_d}\approx 0.48$ the mass ratio of the up/down quark. The second term in the potential is the standard QCD axion potential which ensures $U_{\rm{axion}}(0)=0$, and hence, asymptotic flatness. To gather intuition about the axionic potential, we expand it around its minimum at $\phi =0$,
         \begin{equation}
          U_{\rm{axion}}  \approx \mu _S ^2 \cdot \phi ^2 - \left[ \left( \frac{3B-1}{12}\right) \frac{\hbar\, \mu_S ^2}{f_\alpha ^2} \right] \cdot \phi^4 + \mathcal{O}(\phi ^6)\ .
          \label{E10}
         \end{equation}
   A decrease in the coupling strength $f_\alpha$ implies a decrease in the width of the potential, which is equivalent to an increase in the self-interaction, $f_\alpha \propto 1/\sqrt{\lambda}$. Notice that for the axionic potential, the massive, non-self-interacting configuration is recovered as $f_\alpha \to + \infty$ ~\cite{kaup1968klein,ruffini1969systems}.
   
   The gravity-matter field equations for this case reduce to 
	\begin{align}
         & m' = {4\pi} r^2 \left[N \phi'^{2}+\frac{\omega ^2 \phi ^2}{N \sigma ^2}+U_i \right]\ , \qquad \qquad\sigma' = {8\pi} \sigma r \Big[ \phi'^{2}+\frac{\omega ^2 \phi ^{2}}{N ^2 \sigma ^2}\Big]\ ,\nonumber\\
         & \phi '' = -\frac{2 \phi '}{r}-\frac{N' \phi '}{N}-\frac{\sigma ' \phi '}{\sigma}-\frac{\omega ^2 \phi}{N^2 \sigma ^2}+{\frac{\hat{U_i}}{N}}\phi\ ,
         \label{E11}
	\end{align}
 reminding that $\hat{U_i} \equiv d U_i/d |\Phi |^2$.
%
    \subsection{Numerical strategy and boson star solutions}\label{S2.2}
%
    In order to integrate the set of field equations \eqref{E8} and \eqref{E11} one has to implement suitable boundary conditions at both spatial origin and asymptotic infinity. At the origin, we require that $m (0)=0,\ \sigma (0) =\sigma _0 ,\ \phi (0)=\phi _0 , \ f(0)=f_0$ and $g(0)=0$. In addition, to ensure regularity at the centre, the free functions in the metric and interaction potentials can be expanded around $r = 0$ as 
        \begin{align}
         \label{E12}
         & s=0:~~
	   m  \approx {\frac{4\pi}{3}} \frac{U_i \sigma _0 ^2+\omega ^2\, \phi _0 ^2}{\sigma _0}\, r^3  ,
		\quad \sigma  \approx  \sigma _ 0 + {4\pi}\, \frac{\omega ^2\, \phi _0 ^2}{\sigma _0}\, r^2  , \quad \phi \approx  \phi _0 +\frac{\phi _0}{6} \left( \hat{U_i}-\frac{\omega ^2}{\sigma _0 ^2}\right) r^2  \ , 
		\\
        & s=1:~~m \approx  {\frac{4\pi }{6}}\frac{f_0 ^2\, \mu _P ^2}{\sigma _0 ^2} \, r^3 \ , \quad \sigma  \approx  \sigma _0 + 2\pi\, \frac{ f_0 ^2\, \mu _P ^2}{\sigma _0 }\, r^2 \ ,\quad g \approx -\frac{f_0\, \omega }{3\,\sigma _0 ^2}\, r\ ,  \quad f \approx f_0 +f_0\frac{\mu _P ^2\, \sigma _0 ^2-\omega ^2}{6\, \sigma _0 ^2}\, r^2  \ .
	\end{align}
 At infinity, we impose asymptotic flatness and a finite ADM mass. This translates to the conditions $m(\infty ) =M,\ \sigma (\infty) = 1,$  and $ \phi (\infty) =0=f(\infty)=g(\infty)$. The values of $\sigma _0$ and $M$ are fixed by the numerics, while $\sigma (\infty)$ fixes  the following scaling symmetry of  the system of equations: $\{ \sigma, \omega, f_0 \} \rightarrow \zeta \{ \sigma , \omega, f_0 \}$, with $\zeta >0$. An additional rescaling invariance holds as 
    $
     \{r,m\} =\zeta \{\bar{r},\bar{m}\},\ \{\omega,\mu\}=\frac{1}{\zeta}\{\bar{\omega},\bar{\mu}\},\ \sigma=\bar{\sigma}\ {\rm and}\ \{f,g\}=\{\bar{f},\bar{g}\},
    $
    which imposes invariance of the product $m(r) \mu$ and $\omega /\mu$. Therefore, we can work in units set by the field's mass, \textit{i.e.} $\zeta = \mu ^{-1}$, and set for the scalar and vector mass $\mu _P ^2 = \mu _S ^2 =1$. 
  
  For each value of $\omega /\sigma_0$, the set of coupled ODEs are numerically integrated by means of a Runge-Kutta method with a local error of $10^{-15}$. The boundary conditions are enforced through a shooting strategy on $\phi _0$ (scalar) and $f_0$ (vector) respectively, with a tolerance of $10^{-9}$ set at spatial asymptotic infinity for the scalar/vector decay value, while $m(\infty)\rightarrow M$ and $\sigma(\infty) \rightarrow 1$. To test the numerical solutions we have considered two accuracy tests: the so-called virial identities~\cite{herdeiro2021virial,herdeiro2022deconstructing,oliveira2022convenient}\footnote{A set  of identities obtained from a Derrick-type scaling argument~\cite{derrick1964comments}, which are independent of the equations of motion.}, as well as the difference between the ADM mass (computed as a volume integral of the density \cite{komar1963positive}) and the one obtained from the mass function at infinity. Both tests yield a relative error of $<10^{-5}$. For a more in-depth computation of both expressions, we refer to \cite{herdeiro2021imitation,herdeiro2017asymptotically,herdeiro2020asymptotically}.

\begin{figure}[H]
		   \centering
     	      \begin{picture}(0,0)
		  	 \put(74,186){$U_{\rm{axion}}$ (see \eqref{E9})}
                  \put(74,164){$U_{\rm self}=\frac{\mu_S ^ 2}{2}\phi^2$}
                  \put(74,142){$U_{\rm self}=\frac{\mu_S ^ 2}{2}\phi^2+\lambda \phi^4$}
                  \put(74,120){$V=\frac{\mu_P ^ 2}{2}\textbf{A}^2$}
	   	  \end{picture}
             \includegraphics[scale=1.2]{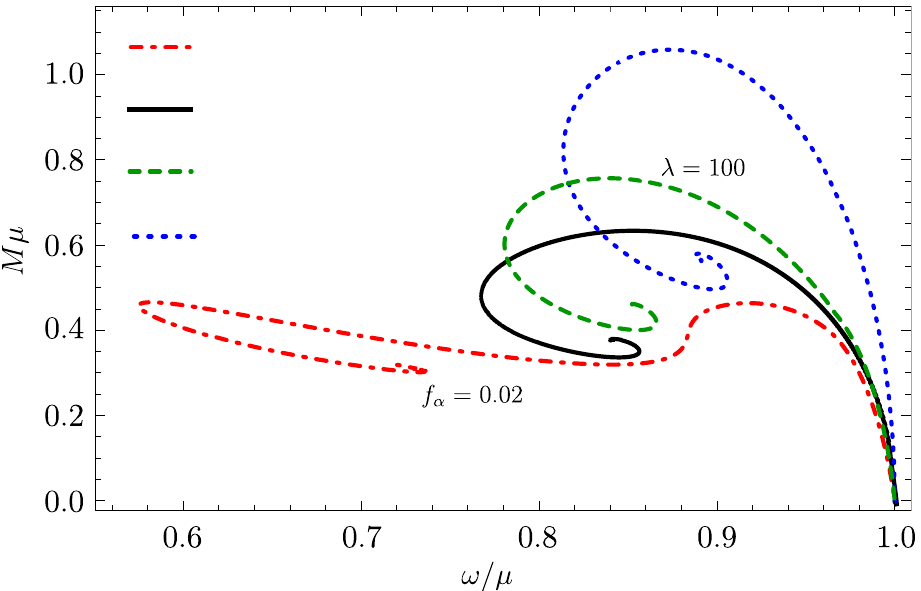}
             \caption{Domain of existence for all our boson star configurations in geometrised units and $\mu =1.0$. $M \mu$ is the total mass of the star in units of $\mu$, while $\omega /\mu$ is a measure of the fundamental energy state of the boson field (see also Section \ref{sec:theory}). The dotted/blue curve corresponds to a star made out of a vector field; the solid/black to a scalar boson star without a self-interaction; the green/dashed to a scalar star with a quartic self-interaction $\lambda=100$; the red/dot-dashed to a scalar star with an axion-like potential with $f_\alpha =0.02$ (see equations \eqref{E9} and \eqref{E10}). At $\omega/\mu = 1$ and $M\mu = 0$, stars tend to have zero density. The mass $M$ ($\omega$) increases (decreases) until it reaches a maximum mass ($M_{\rm{Max}}$) at $\omega_{crit.}$, accompanied with an increase in the star's density. Then the mass decreases until a minimum frequency completing the first branch. Further backbendings and respective branches follow. The first branch up to $M_{\rm{Max}}$ corresponds to the perturbatively stable boson star solutions.}
	 	 \label{F1}
		\end{figure}
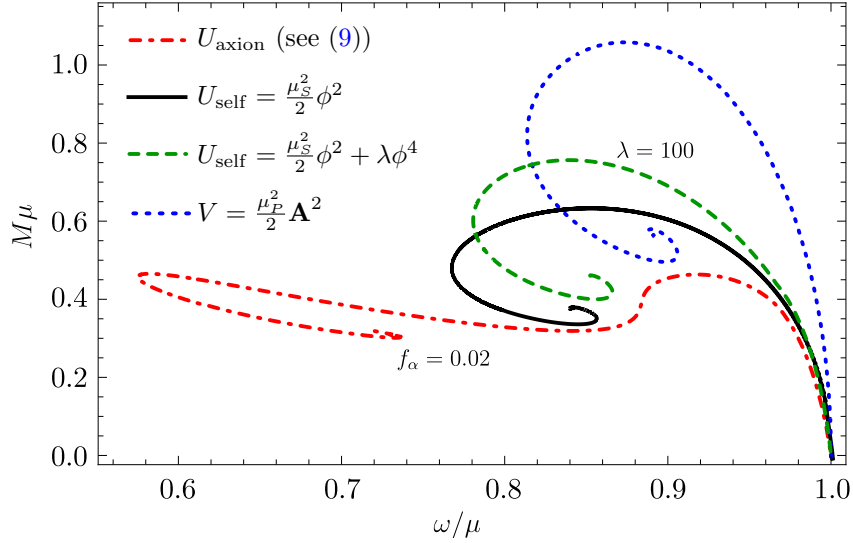

    Iterating our numerical procedure for a sufficiently large set of $\omega /\sigma_0$ points, we can reconstruct what is known as the {\it domain of existence} for our set of models. The result is shown in Figure~\ref{F1}. The figure shows the resulting mass of the boson star against the frequency $\omega$ appearing in the stationary ansatz of equations \eqref{E5}-\eqref{E6}. The figure considers the following cases: a non-self-interacting vector star; a scalar boson star with a quartic self-interaction of $\mu = 1,\ \lambda=100$; a scalar boson star with $\mu = 1, \lambda = 0$, and a scalar boson star with an axion-like potential with $f_\alpha =0.02$. The domain of existence in all models corresponds to a spiral in the $M$ \textit{vs.} $\omega$ diagram starting from $M=0$ for $\omega =\mu $, in which limit the fields become very diluted and the solution trivializes. As one goes inside the spiral, there is an increase of the mass while $\omega$ decreases until the mass attains a maximum $M_{\rm{Max}}$ for  $\omega _{crit.}$. The frequency proceeds to decrease until it reaches a minimum and backbends into a second branch. Further backbendings and branches follow. At the centre of the spiral, the solutions likely tend to a singular solution, however, without ever having a horizon. The first branch up to $M_{\rm{Max}}$ corresponds to the perturbatively stable boson star solutions, which is a common feature for both scalar and vector stars ~\cite{gleiser1989gravitational,gleiser1988stability,lee1989stability}. In the case of the axionic model, an additional stable branch between the first local minimum of the mass and the second local maximum mass exists\footnote{While all the solutions in the first stable branch are energetically stable $Q\mu <M$, only some of the solutions in the second branch are energetically stable (see \cite{herdeiro2021imitation}).}.
%

%
    \section{Orbital dynamics of the two-body system}\label{sec:orbits}
%
   We want to describe the orbit of a luminous star around a dark object as observed by GAIA \cite{el2022sun}. The observed total mass ($M_T$), eccentricity ($e$), period ($T$) and semi-major axis ($a$) respectively read as 
        \begin{equation}
         \{ M_T, e, T, a \} = \{ 11.71 \, M_\odot, \; 0.45, \; 187.77 \, \rm{days}, \; 1.41 \,\rm{AU} \}.
         \label{E14}
        \end{equation}
   The system's total mass consists of a central dark object with mass $M_{\rm{DO}}=9.62\ M_\odot$ and an orbiting luminous star of mass $M_*=0.93 \ M_\odot$. At this stage, it is useful to restore units to facilitate comparison with observations. The metric ansatz \eqref{E5} can be cast in physical units such that $c = 63241\ {\rm AU}\cdot \rm{yr}^{-1},$ $ G = 39.748 \, \rm{AU}^3\cdot M_\odot\cdot \rm{yr}^{-2}$ and $\rm{M_{\rm{Pl}}} = 1.094\times 10^{-38}\ M_\odot\ $. 
    Concerning the orbital dynamics of the luminous star, to simplify the computation we consider the origin of the coordinate systems at the central object. The system's spherical symmetry allows us to restrict the orbit of the luminous star on the equatorial plane around the dark object\textit{, i.e. $\theta =\pi/2$}. There are two conserved quantities associated with the integrals of motion of the orbiting star, namely, the energy $\mathcal{E}$, and the angular momentum $L$ as
    \begin{equation}
\dot{t}= -\frac{\mathcal{E}}{N\sigma ^2\, c^2}\ , \; \; \; \;  \dot{\varphi}=\frac{L}{r^ 2}\ .
    \end{equation}
    The motion of the luminous star in the spacetime background of the central dark object is described by the timelike geodesic equation
        \begin{equation}
         \ddot{x}^\rho+\Gamma ^\rho _{\alpha\beta} \dot{x}^\alpha \dot{x}^\beta = 0\ ,
        \end{equation}     
   with a dot denoting derivative with respect to the affine parameter. The geodesic equation obeys the constraint equation $g^{\alpha \beta} u_{\alpha} u_{\beta} = -c^2$, with $k = c^2$ ($k=0$) for timelike (null) orbits. Expanding it, we get $k = c^2 \cdot \left( N \sigma ^2\, \dot{t}^2 \right) + N^{-1} \dot{r}^2 + r^2 \dot{\varphi}^2$, and after replacing the conserved quantities and solving for the radial velocity, we obtain
        \begin{equation}
         \dot{r}^2 = \left(c^2 -\frac{L^2}{r^2}\right)N+\frac{\mathcal{E}^2}{\sigma ^2 \, c^2}\ .
         \label{E16}
        \end{equation}
   The periastron and apoastron of the orbit are given by $r_p = a\, (1-e)$ and $r_a=a\, (1+e)$ respectively, and we define the useful quantity
        \begin{equation}
         \frac{r_a}{r_p} = \frac{1+e}{1-e} = \xi.
        \end{equation}
   At the periastron and apoastron the radial velocity is zero, \textit{i.e.} $\dot{r}=0$. At these points, equation \eqref{E16} imposes the conditions for the orbit's energy and angular momentum that guarantee the desired orbital properties as a function of $r_a$ and $e$,\footnote{Observe that, for non-vacuum solutions with a matter distribution in orbiting space, one can not guarantee that the orbit has exactly eccentricity $e$ and periastron $r_p$. The previously obtained conditions are only valid for vacuum, \textit{i.e.} the mass is not radius dependent. As we will see in Section ~\ref{sec:CaseII}, the presence of bosonic matter in the orbital region of the star creates non-trivial distortions of the orbit.}  
        \begin{equation}
         \mathcal{E}^2 =  -k\cdot \frac{ (\xi +1)\, \sigma^3 \left(c^2\, r_a - 2\, G\, m\right) \left(c^2\, r_a-2\, G\, \xi\,  m \right)}{r_a \Big[c^2 (\xi +1)\, r_a-2\, G \left(\xi ^2+\xi +1\right) m\Big]}\ ,\qquad
         L^2 = - \frac{2\, c^2\, G\, m\, r_{a}^2}{c^2\, r_{a}(\xi + 1) - 2\, G\, m\, (1 + \xi + \xi^2)}\ .
         \label{E18}
        \end{equation}
We remind that in our numerical construction of the boson star solutions, we used geometrised units with $\mu=1$. To use the previous solutions as a proper dark object mimicker we have to impose the proper scale. Following previous works one notices that \cite{schunck2003general,colpi1986boson,schunck2000boson,ho1999maximum}
        \begin{equation} \label{eq:M_DO}
         M_{\rm{DO}} = M\, \frac{M_{\rm{Pl}}^2}{\mu}\ ,
        \end{equation}
    with $M$ the mass of the boson star obtained in the previous Section~\ref{sec:theory} in geometrised units with $\mu=1$, and $M_{\rm{DO}}$ the observed mass of the central dark object in units of $M_\odot$. Since both $M_{\rm{DO}}$ and $M$ are known, the previous relation allows one to recover the particle's mass $\mu$ and the correct rescaling of all quantities. The radial coordinate scales as
        \begin{equation}
         R = \frac{G M_{\rm{DO}}}{M \, c^2}\ r \ .
        \end{equation}
      Due to the lack of a hard surface boson stars do not have a well-defined radius. For the ``radius`` of the latter, let us consider the areal radius of a spherical surface within which $99\% $ of all the mass is included ($R_{*}$). The latter defines the boson star compactness as $\mathcal{C}\equiv 2M_{99}/R_{*}$. The boson star compactness is always smaller than unity, becoming unity for black holes. 
      
      We can now notice the existence of two kinds of orbits. \textbf{Case I} is characterised by $r_p>R_{*}$, and corresponds to the case where the luminous star never passes through a region of the spacetime with significant matter density. In a similar way, \textbf{Case II} correspond to $r_p<R_{*}$, and in this case, the orbiting star has at least part of its orbit crossing a relevant matter density region.

%
    \section{Case I: $r_p > R_{*}$} \label{sec:CaseI}
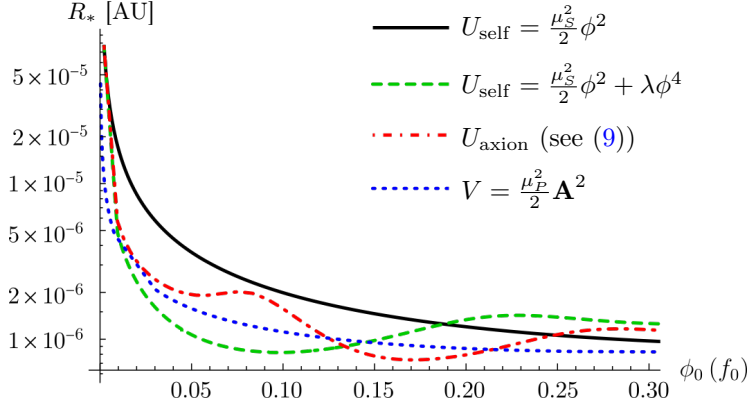
\begin{figure}[H]
	   \centering
            \begin{picture}(0,0)
		  	 \put(174,95){$U_{\rm{axion}}$ (see \eqref{E9})}
                  \put(174,137){$U_{\rm self}=\frac{\mu_S ^ 2}{2}\phi^2$}
                  \put(174,115){$U_{\rm self}=\frac{\mu_S ^ 2}{2}\phi^2+\lambda \phi^4$}
                  \put(174,75){$V=\frac{\mu_P ^ 2}{2}\textbf{A}^2$}
                  \put(25,143){\small$R_*$ [AU]}
	    \end{picture}
          \includegraphics[scale=1.1]{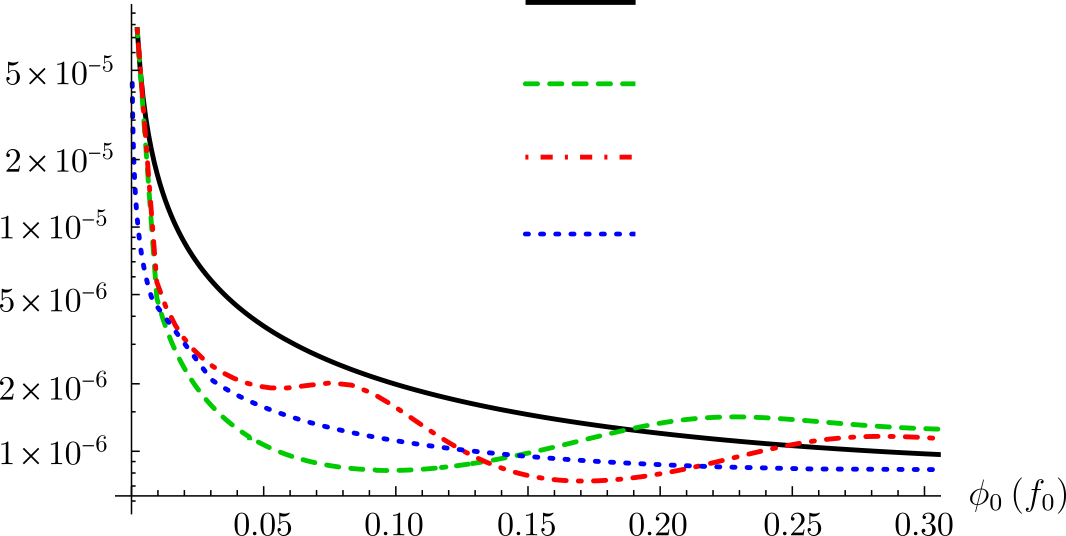}\\
          \caption{Radius ($R_{*}$) in astrophysical units containing $99\%$ of the total mass of a boson star as a function of the respective field amplitude at the origin -- $\phi_0$ for the scalar field and $f_0$ for the vector field. The dotted/blue curve corresponds to a star built out of a vector; the solid/black to a scalar boson star without a self-interaction ($\lambda =0$); the green/dashed to a scalar star with a quartic self-interaction $\lambda = 100$; and the red/dot-dashed to a scalar star with an axion-like potential $f_\alpha = 0.02$ (see equations \eqref{E9} and \eqref{E10}). The increase in the field's central amplitude corresponds to going inside the domain of the existence spiral from the right (see Figure~\ref{F1}). The leading term for small values of the field's amplitude is the mass term, while the self-interaction/axion potential is only relevant for larger values of the field's amplitude. In all cases, the maximum of $R_{*}$ occurs in the region dominated by the mass term, always far smaller than the orbital radii reported by GAIA \cite{el2022sun}.}
	   \label{F2}
	\end{figure}

%
Let us start by imposing the observational parameters of the GAIA observation \eqref{E14}.
 In this case, $r_p\approx 0.8$ $\rm{AU}$, while the radius of a scalar boson star without a quartic interaction, \textit{i.e.} $\lambda=0$, is $R_{*} \approx 6.7 \times 10^{-6}$ $\rm{AU}$ (around $100\ \rm{km}$) at maximal mass configuration $M_{\rm{Max}}$ \cite{schunck2003general} (see Section \ref{S2.2}). This corresponds to a ratio of 
 \begin{equation}
 R_{*}/r_{p} \approx 6.67 \times 10^{-7}. 
 \end{equation}
 Consequently, the luminous star never experiences a relevant matter density distribution. The luminous star is so far from the central dark object that, as seen in Figure~\ref{F2}, it is impossible to distinguish the dark object's nature. 
 In fact, looking at both the computed eccentricity and orbital period (our two tests), one can see  that independently of the boson star's model, the latter coincide with the values obtained from the GAIA observation \eqref{E14}. Both eccentricity and period, $(e,\, T)$, are highly stable across the domain of existence and between models. 
 
    However, one possibility remains for the luminous star's orbit to be altered by the matter distribution associated with the dark object. As one goes outside the domain of existence spiral (see Figure~\ref{F1}), from the maximum mass $M_{\rm{Max}}$ to the highly diluted regime ($\omega \to \mu$), one observes a decrease in the compactness of the solution, predominantly driven by an increase in the radius $R_{*}$ (see Figure~\ref{F2}). 
    
    In addition, as already stated in Section~\ref{S2.2}, the region between the maximal mass $M_{\rm{Max}}$ and $\omega =1.0$ corresponds to the boson star's stable branch \cite{herdeiro2021imitation,gleiser1988stability,gleiser1989gravitational,brito2016proca}. An additional stable branch that goes from the second maximum of the mass (the absolute maximum) to some value of $\omega < \omega _{crit.}$ \cite{herdeiro2021imitation} exists in the axion-like case.

    Since $R_*$ at $M_{\rm{Max}}$ scales as $\sim \sqrt{\lambda}$ for self-interacting scalar boson star, with non-self-interacting solutions recovered for $\lambda =1$ \cite{schunck2003general}, one could ask if the presence of a positive self-interaction is enough to increase $R_*$ such that $R_* \sim r_p$. However, the self-interacting scalar boson star solution has a higher mass at the maximum of the existence diagram (see Figure \ref{F1}), and the resulting configuration in astrophysical units will have a smaller radius than the non-self-interacting solution. In addition, the quartic self-interaction tends to be relevant for solutions which lie inside the spiralling part of the domain of existence. The latter solutions are simultaneously unstable and not significantly altered for solutions we can numerically obtain. In particular, when $\lambda$ is too large, there is an increased numerical difficulty due to the couplings $\mu$ and $\lambda$ being separated by many orders of magnitude. In addition, the boson star tends to a self-interacting system instead of a self-gravitating one, which is forbidden for positive quartic self-interactions \cite{derrick1964comments,herdeiro2021virial}.

    The same behaviour occurs for the vector star solutions as in the case of a scalar star with $\lambda=0$. Following the same line of thought as with the scalar boson stars, one could add a quartic self-interaction potential. However, as shown in \cite{minamitsuji2018vector}, no significant value of the quartic self-interaction constant would be achieved\footnote{Recent works~\cite{clough2022problem} raised the existence of ghost instabilities in quartic self-interacting vector star models.}.

    Finally, in the case of the axion-like boson star, in neither branch of the existence diagram was possible to obtain a boson star with $R_* \sim r_p$. Hence, no observable deviation from the measured orbital properties \eqref{E14} was found to exist within the explored parameters. However, it could be the case that if one decreases the coupling constant $f_\alpha$ further below $f_\alpha = 0.02$, that the negative self-interaction characteristic of the potential $U_{\rm axion}$ in \eqref{E9} would lead to a more diluted boson star that tends to have a size $R_*$ larger than a corresponding non-interacting boson star. 

     The results of our analysis for the Case I are shown on the top left plots of Figure~\ref{fig:Orbits-scalar} (scalar boson star) and Figure~\ref{fig:Orbits-vector} (vector boson star). The orbital properties of the luminous star, namely, eccentricity, period and semi-major axis, reproduce the observed values from GAIA. We should notice that for visualisation purposes we have rescaled the size of the boson star with respect to the orbital scale $r_a$, so that $r_a /R_{*} = 10$, while in the previous section $r_a /R_{*} \sim 10 ^{6}$. The latter choice would just make the inner boson star a simple dot in the figure. 
     
To summarise, we showed in this section that the measured orbital parameters are consistent with the scenario that the dark object detected by GAIA~\cite{el2022sun} is a boson star of either scalar or vector nature for natural choices of the respective theory spaces.

\begin{table}[H]
 \centering
 \caption{Characteristics of the boson star model belonging to Case I (see Section \ref{sec:CaseI}), which gives configurations in agreement with GAIA's observation. All boson star configurations below correspond to the maximum ($M_{\rm{Max}}$) of the respective curve on the existence diagram (see Figure \ref{F1}). The total physical mass is fixed as $M_{T}=11.77\,M_{\odot}$ according to GAIA's observation. Notice that for the scalar/vector particle's mass ($\mu$) the boson star is built of, we give only the respective magnitude. }\label{T1}
 \vspace{3mm}
\begin{tabular}{ |p{3.5cm}||p{3cm}|p{2.5cm}|p{2.5cm}|p{1.6cm}| }
 \hline
 \multicolumn{5}{|c|}{\bf Boson star characteristics - Case I} \\
 \hline
    Model & Potential & Boson mass ($\mu$) at $M_{\rm{Max}}$ [${\rm GeV/c^2}$] & Radius $R_{*}$ at $M_{\rm{Max}}$ [km] & Interaction coupling ($\lambda$) \\
 \hline
    Scalar, non-interacting & $U_{\rm self}=\frac{\mu_S ^ 2}{2}\phi^2$ & $10^{-20}$ & $134$ & $\lambda=0$ \\
    Scalar, self-interacting & $U_{\rm self}=\frac{\mu_S ^ 2}{2}\phi^2+\lambda \phi^4 $ & $ 10^{-20}$ & $87$ & $\lambda= 100$\\
    Scalar, axion-like & $U_{\rm{axion}}$ (see \eqref{E9}) & $10^{-19} $ & $317$ & $f_\alpha = 0.02$ \\
    Vector & $ V=\frac{\mu_P ^ 2}{2}\textbf{A}^2$ & $ 10^{-20}$ & $124$ & $0$\\
 \hline
\end{tabular}
\end{table}

    \section{Case \textbf{II}: $r_p < R_{*}$} \label{sec:CaseII}
%
    In the previous section, we showed that one can construct fairly natural boson star configurations where the luminous star is always very far from the matter distribution generating the central gravitational potential. As a result, the orbit of the luminous star is practically indistinguishable from the one resulting from a black hole as a dark object. However, here we would like to entertain the possibility that the boson star could be diluted enough to have the orbiting star inside the cloud ($r_p<R_{*}$). While a thorough study of timelike  orbits in a boson star background is out of the scope of this work, we would like to qualitatively show how different these can be from a point-like source. As we will see, the trajectory of the luminous star for this case is drastically different from Case I. We will also discuss later what potentials could lead to such diluted boson star configurations.  
    
\begin{figure}[H]
	   \centering
          \includegraphics[scale=0.25]{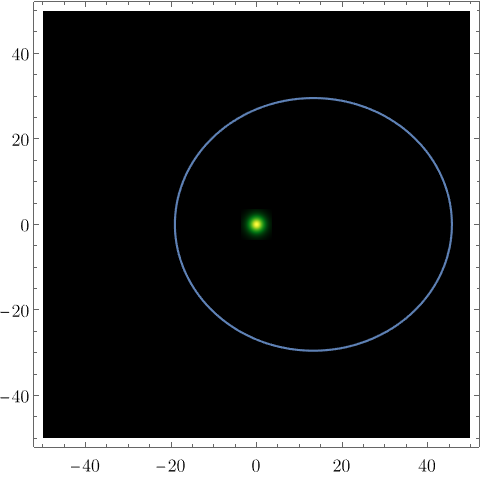}
          \includegraphics[scale=0.25]{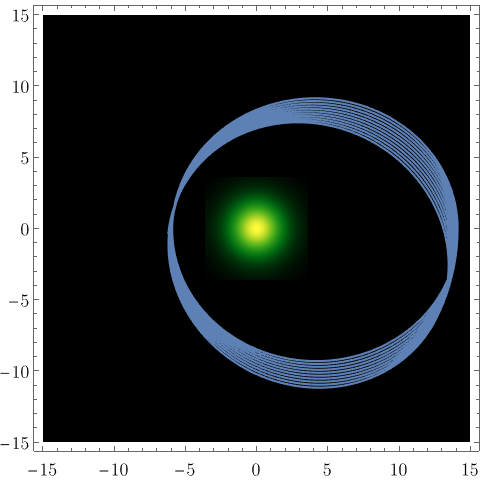}
          \includegraphics[scale=0.465]{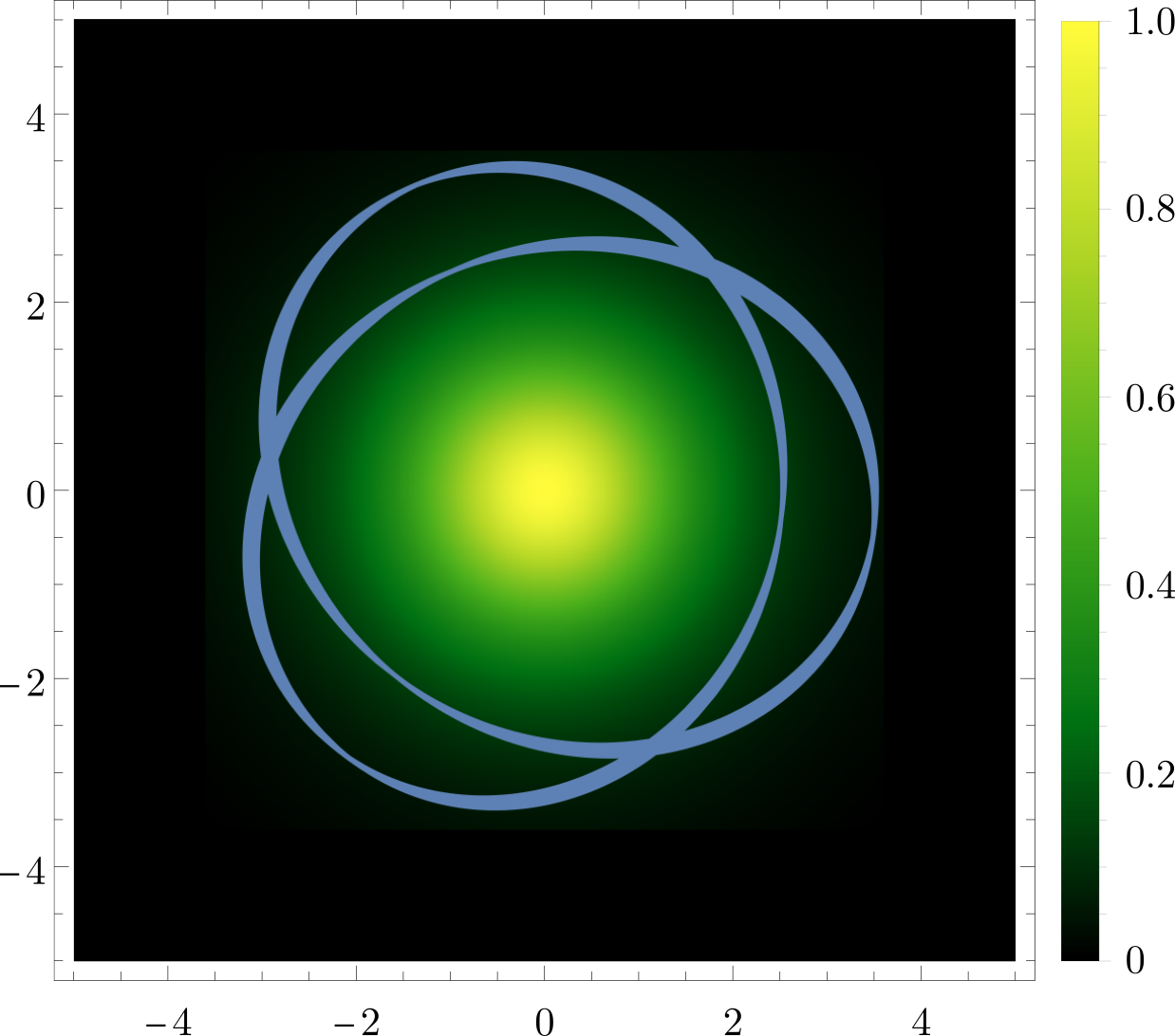}\\
          \includegraphics[scale=0.25]{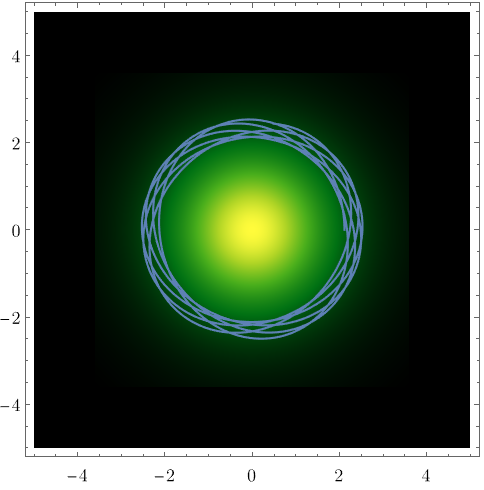}
          \includegraphics[scale=0.25]{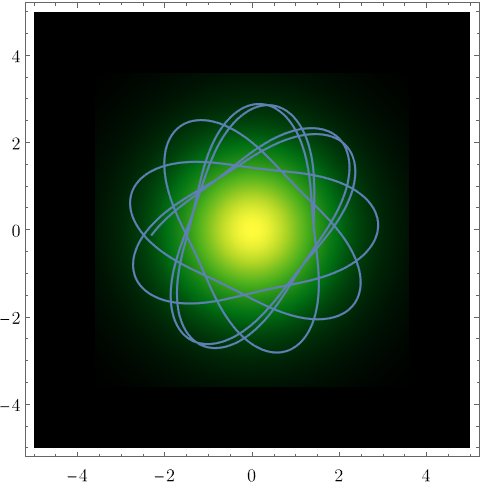}
          \includegraphics[scale=0.47]{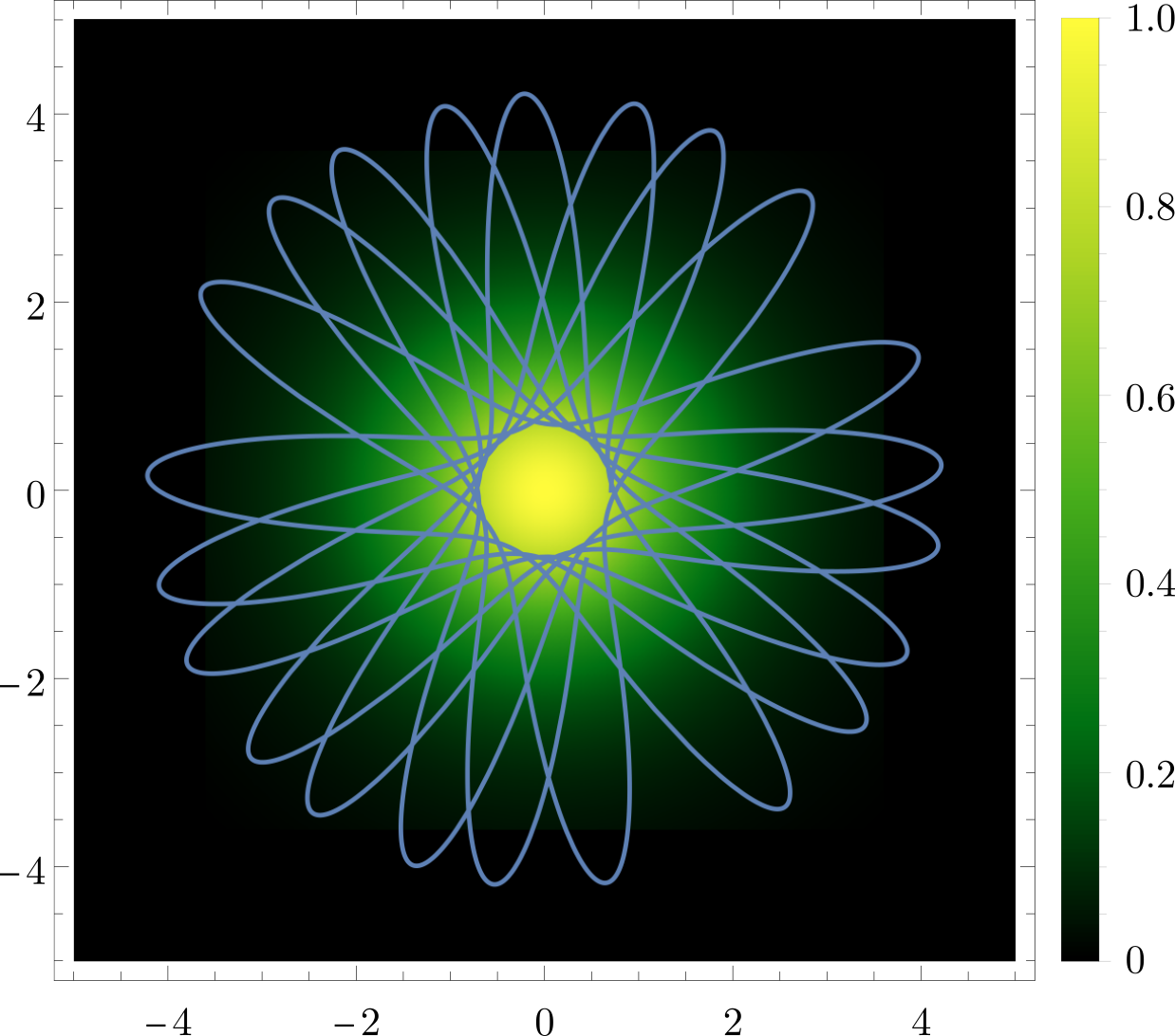}\\
          \includegraphics[scale=0.25]{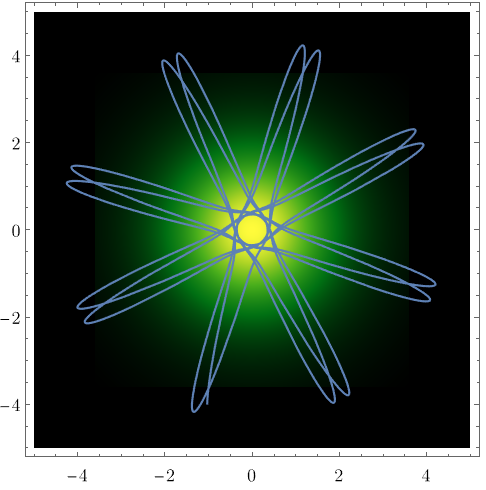}
          \includegraphics[scale=0.47]{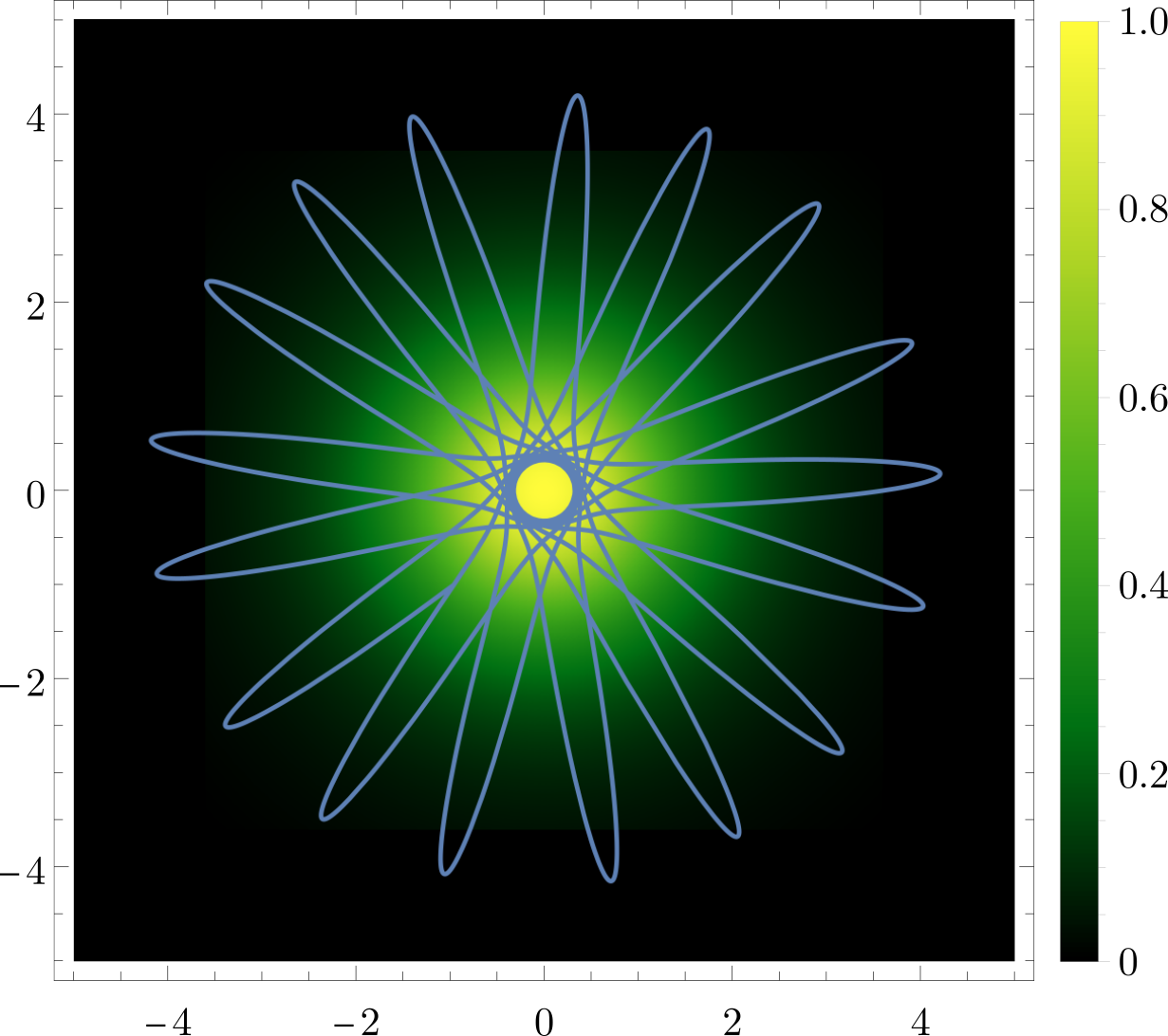}
          \caption{Geodesic motion of a point particle (luminous star) around a scalar boson star. All boson star configurations correspond to the following theory space choice: $\omega = 0.8$, $M = 0.5$, $\lambda=0$ and $R_{*} = 4.6$ (see Section \ref{sec:theory}). The individual plots correspond to the following initial conditions/choices for the apoastron ($r_a$), boson star size ($R_*$) and eccentricity ($e$): (Top left) $r_a/R_{*} =10.0 $ and $e = 0.45$; (Top middle) $r_a/R_{*} =3.1 $ and $e = 0.4$; (Top right) $r_a/R_{*} =0.8 $ and $e = 0.2$; (Middle left) $r_a/R_{*} =0.5 $ and $e = 0.1$; (Middle middle) $r_a/R_{*} =0.3 $ and $e = 0.3$; (Middle right) $r_a/R_{*} =0.2$ and $e = 0.7$; (Bottom left) $r_a/R_{*} =0.1 $ and $e = 0.9$; (Bottom right) $r_a/R_{*} =0.07 $ and $e = 0.9$. In the background, we plot the amplitude of the scalar field normalized to its amplitude at the origin (maximum), $\phi /\phi_0$. Notice that a lack of precession and the correct orbital eccentricity only occurs in the top left case, which is also the one exactly reproducing the GAIA observation (see Case I of Section \ref{sec:CaseI}). In the other plots, precession is induced due to the orbit of the luminous star crossing the bosonic matter distribution (see  Case II of Section \ref{sec:CaseII}). For Case II, the eccentricity of the orbits does not correspond to the one provided as an initial condition. }
	   \label{fig:Orbits-scalar}
	\end{figure}

\begin{figure}[H]
		 \centering
          \includegraphics[scale=0.25]{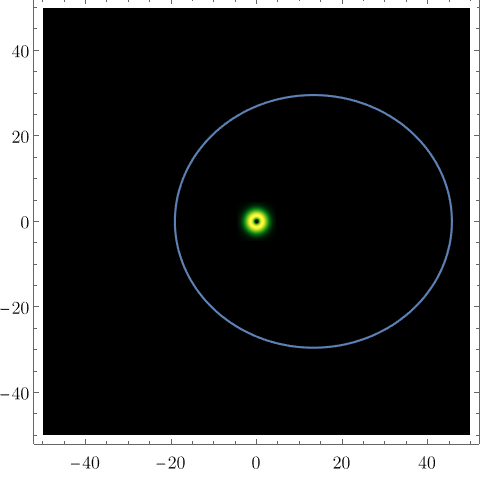}
          \includegraphics[scale=0.25]{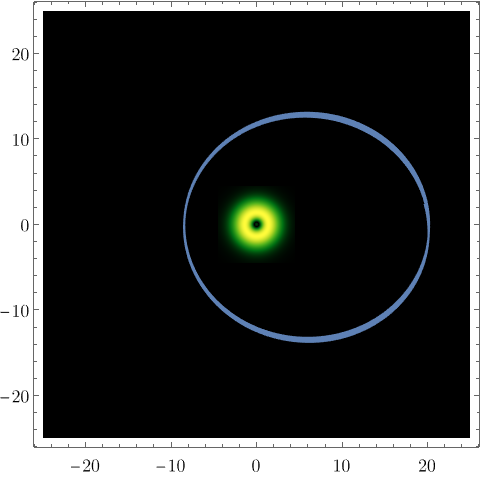}
          \includegraphics[scale=0.465]{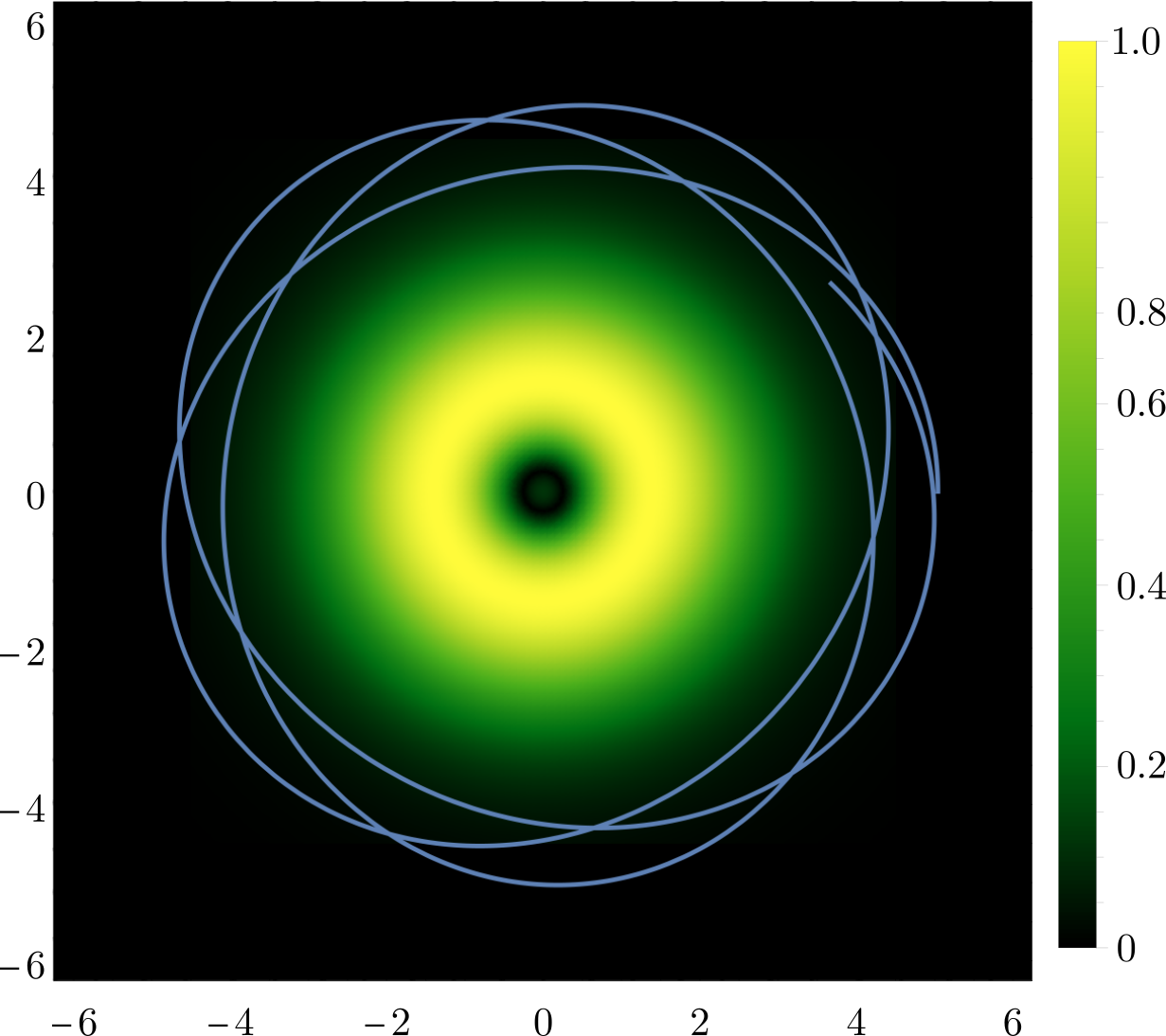}\\
          \includegraphics[scale=0.25]{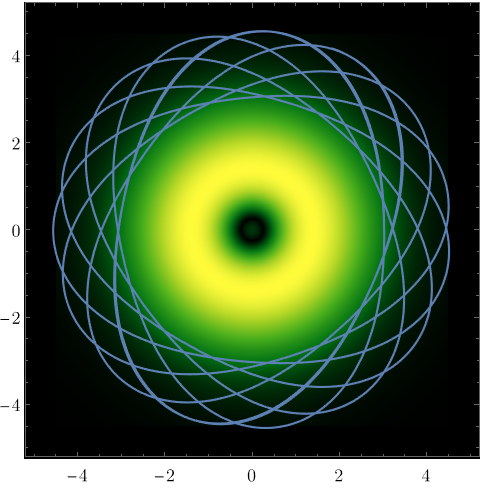}
          \includegraphics[scale=0.25]{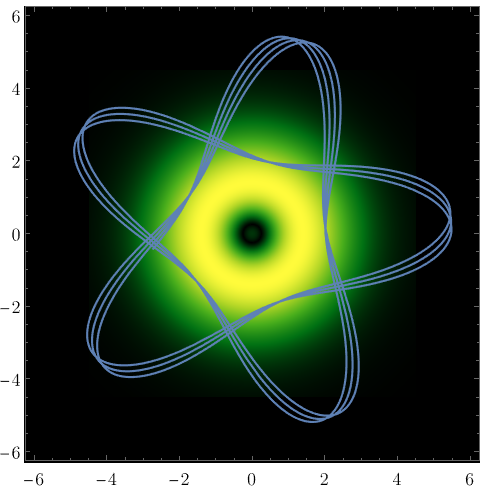}
          \includegraphics[scale=0.47]{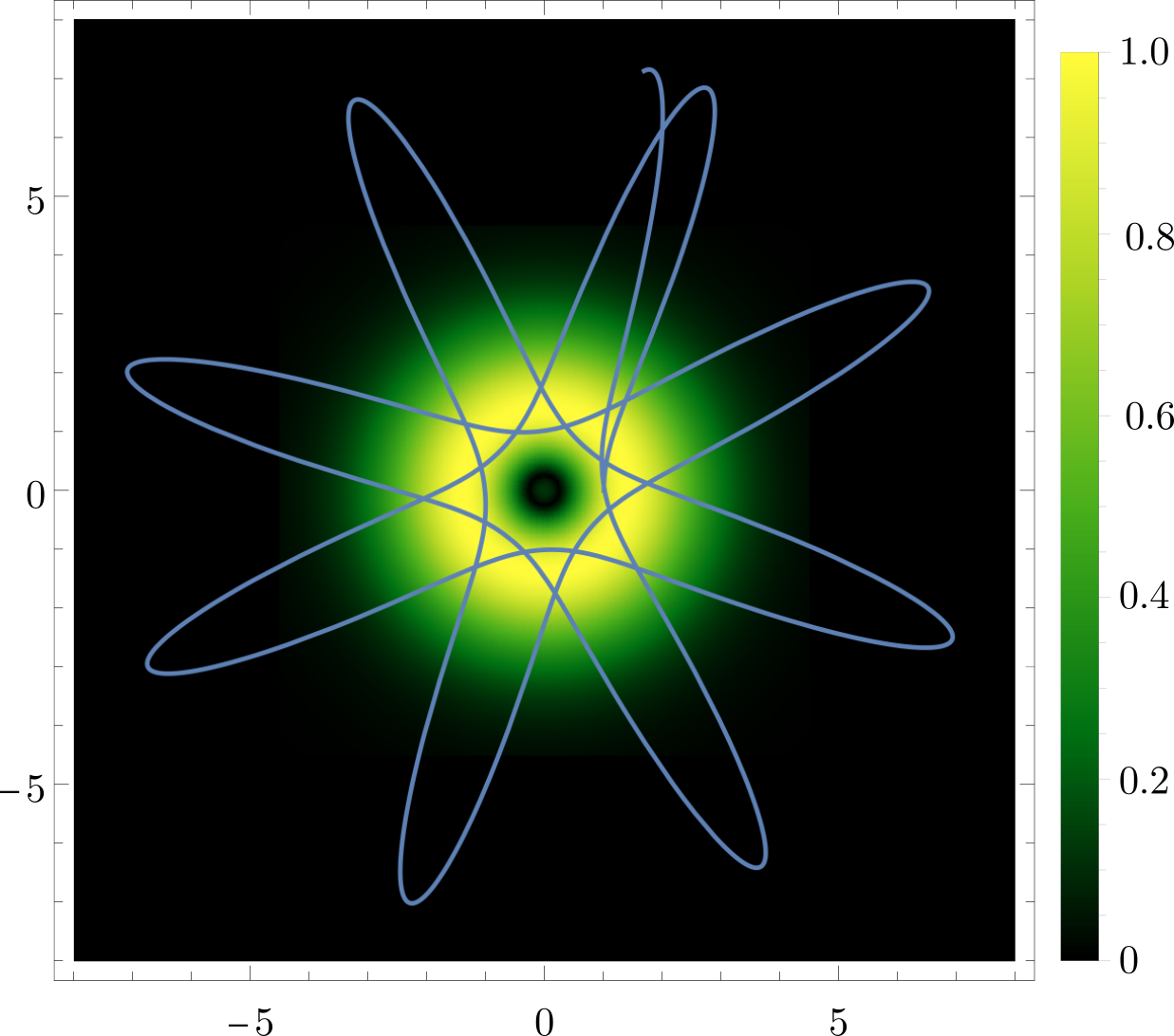}\\
          \includegraphics[scale=0.25]{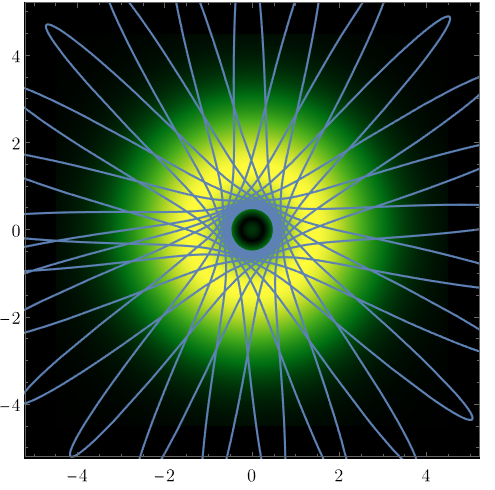}
          \includegraphics[scale=0.47]{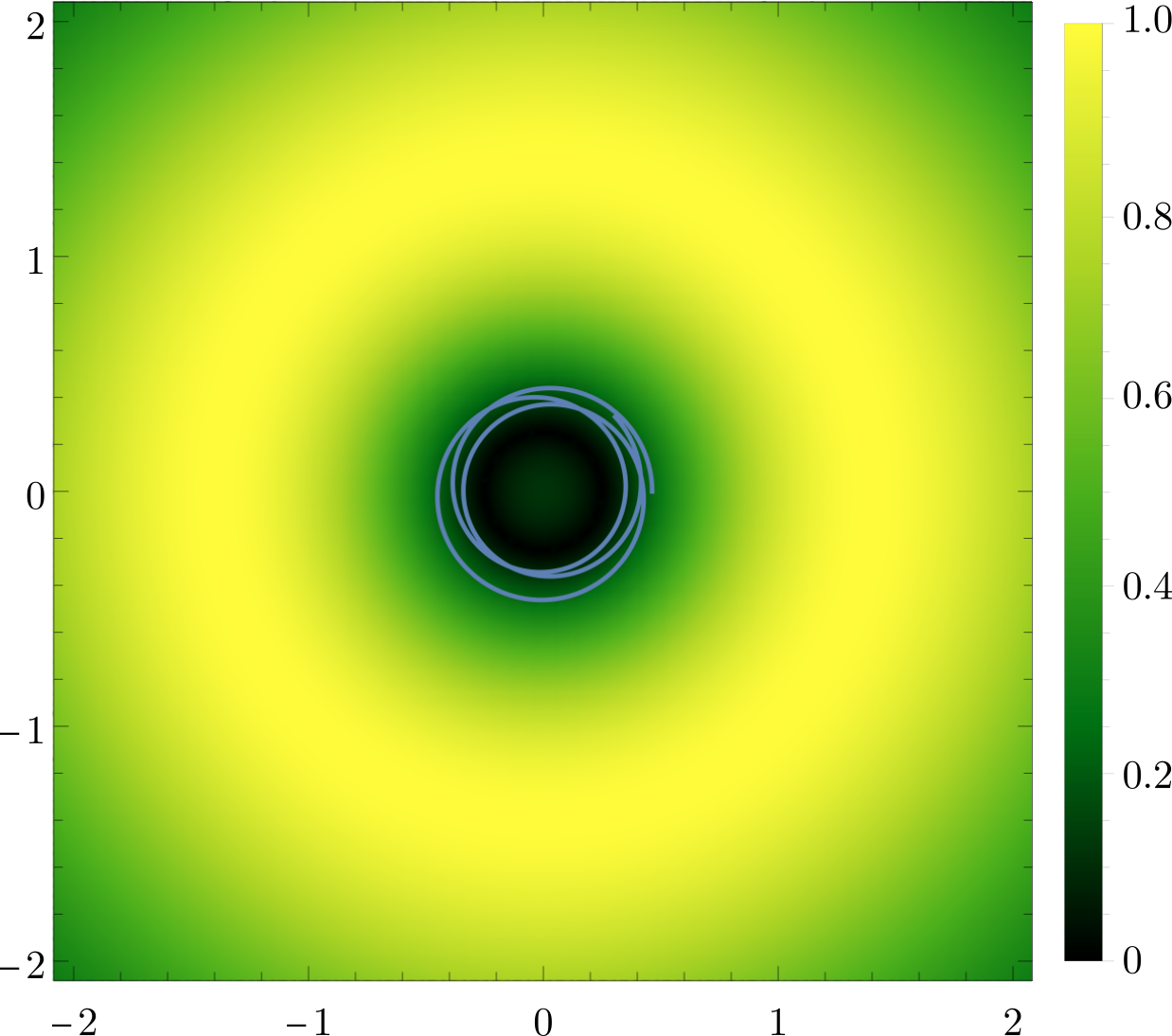}
          \caption{Same as Figure \ref{fig:Orbits-scalar}, but for the geodesic motion of a point mass orbiting a vector boson star. All shown boson star configurations correspond to the following theory space choice: $\omega = 0.8$, $M = 0.8$ and $R_{*} = 6.5$ (see Section \ref{sec:theory}). The individual plots correspond to the following initial conditions/choices for the apoastron ($r_a$), boson star size ($R_*$) and eccentricity ($e$): (Top left) $r_a/R_{*} =10.0$ and $e = 0.45$; (Top middle) $r_a/R_{*} =3.1$ and $e = 0.4$; (Top right) $r_a/R_{*} =0.8 $ and $e = 0.1$; (Middle left) $r_a/R_{*} =0.5 $ and $e = 0.2$; (Middle middle) $r_a/R_{*} =0.3 $ and $e = 0.5$; (Middle right) $r_a/R_{*} =0.2 $ and $e = 0.8$; (Bottom left) $r_a/R_{*} =0.1 $ and $e = 0.9$; (Bottom right) $r_a/R_{*} =0.07 $ and $e = 0.2$. In the background is plotted the vector field density amplitude $|\textbf{A}^2|$ of the vector field normalized to its maximum amplitude, $\frac{|\textbf{A}^2|}{{\rm Max} \ \textbf{A}^2}$. We remind that a lack of precession and the correct orbital eccentricity only occurs in the top left case, which is the one exactly reproducing the GAIA observation (see Case I of Section \ref{sec:CaseI}). In the other plots, precession is induced due to the orbit of the luminous star crossing the bosonic matter distribution (see Case II of Section \ref{sec:CaseII}). For Case II, the eccentricity of the orbits does not correspond to the one provided as an initial condition.  }
	 	\label{fig:Orbits-vector}
		\end{figure}

    The simplest way to bring the luminous star within the bosonic cloud is to modify the ratio $r_{a}/R_{*}$, where $R_{*}$ is the radius of the boson star at $99\%$ of its mass. We choose to consider a rescaling of the radial coordinate in such a way that the semi-major axis of the orbit ($a$) is of the order of the boson star $99\%$ radii, $a\sim R_{*}$. Clearly, this will not correspond anymore to the value observed by GAIA, but our goal in this section is to rather understand the qualitative characteristics of such orbits. The results are shown in Figure~\ref{fig:Orbits-scalar} for a scalar star, and Figure~\ref{fig:Orbits-vector} for a vector star, respectively. For each orbit of this case, we provide as initial conditions the apoastron, $r_{a}$, and the eccentricity $e$. Although $r_{a}$ retains its value as the system is let to evolve, this does not happen for the eccentricity which settles down to a different value compared to its initial condition. Observe that as one moves $r_{a}$ deeper within the bosonic cloud, an increase in the eccentricities is observed, accompanied by sufficiently large orbital precession. In the present case, the mass function $m(r)$ changes with the radial position in the region of the orbit. The existence of a matter field in the orbit of the luminous star implies a potential well that is very sharp at the origin and then slowly increases as $r$ increases. This in principle leads to a trajectory with a large range of radii to keep the same orbital energy. 
    
    We can get a qualitative insight for the precession of these orbits if we assume that the star crosses only a thin outer shell of bosonic matter. This allows us to use approximately the relation for precession around a Schwarzchild spacetime as derived in GR, $\delta \varphi \sim G M_{99}/(r_a (1-e^2))$, with $\varphi$ the precession angle, and we assumed an elliptic Keplerian orbit. As one can see, an increase in eccentricity ($e$) will lead to an increase in the amount of precession. Obviously, the validity of this relation breaks down as long as the boson star's mass cannot be considered constant anymore.  
     While for both scalar and vector stars the orbiting body inside the matter cloud has an eccentric and precessing orbit, for the case of a vector star, due to the presence of a maximum of the vector field amplitude $\textbf{A}^2$, a region of the spacetime exists where the orbiting star exhibits small eccentricities (see Figure~\ref{fig:Orbits-vector} bottom right).       
    %
%
           
 The large eccentricities and high orbital precession, which occurs when the luminous star crosses the matter distribution, create a distinctive behaviour compared to the Case I explored in the previous section. {\it This seems to indicate that:} {\bf i)} GAIA~\cite{el2022sun} can not be reproduced by a boson star configuration that encompasses the orbit of the star; {\bf ii)} there is a distinct, observationally-testable behaviour that can be used to probe the nature of the dark object. Indeed, our predicted orbital dynamics for Case II imply that future astrometric observations of similar systems could provide an opportunity to distinguish between a scalar and a vector boson star (or even more exotic fields), and their respective interactions. A detailed prediction of such orbital dynamics goes beyond the scope of our current project, and we leave this for future work, however, our analysis points to the following statement. For a potential with a mass term, $\mu > 0$, the addition of self-interactions with positive couplings will tend to make the star smaller, \textit{i.e.} interactions $\sim c_{2n} \phi^{2n}$ with $c_{2n}>0$ and $n \geqslant 2$. The opposite situation will occur for negative couplings $c_{2n} < 0$, provided they dominate the potential over the mass term. Such a situation could occur for the axion-like potential of equation \eqref{E9} with a sufficiently small coupling $f_\alpha \ll 1$, which will, in turn, lead to a sufficiently negative self-interaction. The latter solutions require a thorough investigation of their own due to non-trivial features compared to the other cases studied in this work. A first step in constructing the respective existence diagram for such a potential has already been made in \cite{guerra2019axion}.
           
    \section{Summary and discussion} \label{sec:summary}
%
    Recently, GAIA observed a Sun-like star orbiting a dark central object, at a distance of about $1.4 \, \rm{AU}$, and a particularly high orbital period of $187.8$ days \cite{el2022sun}. As we explained in the Introduction (Section~\ref{Intro}), identifying the central dark object with a black hole is plausible, but rather challenging to support within the context of binary evolution given the characteristics of the binary system. Here, we investigated the possibility that the central dark object is a boson star, showing that a boson star can explain the observations for fairly reasonable choices of the theory-space parameters.
\\

    For our analysis, we considered two distinct types of boson stars, namely, scalar and vector ones, respectively made of scalar and vector boson particles. We solved the coupled gravity-matter equations to construct the boson star models, which provided the gravitational potential of the dark central object. Our results for the scaling of the boson stars' masses and radii are shown in Figures \ref{F1} and \ref{F2} respectively. 
\\

    We proceeded with the computation of timelike orbits of a point mass (luminous star) in the gravitational potential of the constructed boson stars. The results are shown in Figures \ref{fig:Orbits-scalar} and \ref{fig:Orbits-vector}.  In this regard, we considered the case where the luminous star does not cross the bosonic matter distribution (Case I, Section \ref{sec:CaseI}), and a case where it does (Case II, Section \ref{sec:CaseII}). 
    In \textbf{Case I}, the constructed boson stars exhibit radii which are much smaller than the closest approximation of the orbiting star, \textit{i.e.} $r_p>R_* $, and the matter density present in the space where the luminous star passes through is negligible. As we showed,  one can construct boson star configurations for fairly natural choices of theory-space parameters, for which the luminous star reproduces exactly the observed GAIA orbit, as if the star was orbiting around a central black hole. 
    In \textbf{Case II}, the boson star has a radius close to the size of the orbiting star's closest approximation, \textit{i.e.} $R_* \sim r_p$. As shown in Figures~\ref{fig:Orbits-scalar} and \ref{fig:Orbits-vector}, a non-negligible matter field in the orbit of the luminous star modulates the gravitational potential so that the orbiting star gains large eccentricities and precessions. The absence of an observed precession by GAIA rules out the presence of a matter field in the trajectory of the star, or the latter has to be negligibly small. What is more, from Figure~\ref{F2} and Table~\ref{T1} one notices that the driving term in the solution for the radius of the boson star comes from the mass term in the potential, \textit{i.e.} the term $\propto \phi ^2$. The addition of a quartic self-interaction with positive coupling ($\lambda > 0$) on top of the mass term in the potential makes the corresponding boson star smaller in size. We expect this behaviour to hold for higher-order interactions in the potential, that is, interactions of the form $\sim c_{2n} \phi^{2n}$ with $c_{2n}>0$ and $n\geqslant 2$. On the other hand, for a potential which is dominated by a negative self-interaction, the boson star can be larger than the orbital radius leading to precession and high eccentricities. Such potentials are ruled out by GAIA's observation. We remind that a potential with a negative self-interaction can arise in our axion-like potential given in equation \eqref{E9} with coupling constant $f_\alpha \ll 1$.
\\

     Our analysis highlights the potential of high-precision astrometric observations to test the idea of boson stars and their phenomenological imprints against those of black holes. Future astrometric observations will further offer the possibility to distinguish amongst different boson star scenarios, opening up an exciting window into the search for elusive particles beyond the standard model.
    \section*{Acknowledgments} \label{Akn}
A. Pombo is supported by the Czech Grant Agency (GA\^CR) under the grant number 21-16583M. I.D.S acknowledges support from the Czech Academy of Sciences under the project number LQ100102101. I.D.S is thankful to Benoit Famay for a useful discussion on the GAIA observations.

%

\bibliographystyle{alpha}
\bibliography{sample}

\end{document}